\documentclass[preprint,10pt]{article}
\usepackage{cite}
\usepackage{amsmath,amssymb,amsfonts}
\usepackage{listings}
\usepackage{algorithm,algpseudocode}
\usepackage{graphicx}
\usepackage{textcomp}
\usepackage{authblk}
\usepackage[math]{easyeqn}

\newcommand{\mybm}[1]    {\mbox{\boldmath{$#1$}}}

\newcommand{\boldo}    {{\mybm 0}}

\newcommand{\boldu}    {{\mybm u}}

\newcommand{\boldtau}  {{\mybm \tau}}

\newcommand{\defor}    {{\mybm \varepsilon}}

\newcommand\mnewcommand[1]{%
\let#1\relax \newcommand#1 }

\mnewcommand{\diver}{\mnabla \cdot}
\mnewcommand{\vel}{\mathbf{u}}
\mnewcommand{\velhg}{\vel_h}
\mnewcommand{\preshg}{{p}_h}
\mnewcommand{\dim}{3}
\mnewcommand{\ud}{d}
\mnewcommand{\vort}{\mathbf{w}}
\mnewcommand{\rot}[1]{\mnabla \times #1}
\mnewcommand{\selfinnerprod}[1]{\innerprod{#1}{#1}}
\mnewcommand{\innerprod}[2]{< #1 , #2 >}
\mnewcommand{\lapl}{\Delta}
\mnewcommand{\Nx}{N_1}
\mnewcommand{\Ny}{N_2}
\mnewcommand{\Nz}{N_3}
\mnewcommand{\Nm}{N_m}
\mnewcommand{\Ns}{N_s}
\mnewcommand{\complconj}[1]{#1^{*}}
\mnewcommand{\step}{\Delta}
\mnewcommand{\dt}{\step t}
\mnewcommand{\traspose}{^{*}}
\mnewcommand{\avg}[1]{\overline{#1}}
\mnewcommand{\mcdot}{\bcdot}
\mnewcommand{\mnabla}{\nabla}
\mnewcommand{\real}{\mathbb{R}}
\mnewcommand{\complex}{\mathbb{C}}

\mnewcommand{\matvec}[1]{\mathbf{#1}}
 \mnewcommand{\mA}{\matvec{A}}
\mnewcommand{\mx}{\matvec{x}}
\mnewcommand{\mb}{\matvec{b}}

\providecommand\bnabla{\boldsymbol{\nabla}}

\mnewcommand{\mnabla}{\bnabla}
\mnewcommand{\lapl}{\mnabla^2}
\mnewcommand{\diver}{\mnabla \cdot}
\mnewcommand{\vel}{\mathbf{u}}
\mnewcommand{\vels}{u_s}
\mnewcommand{\gradvel}{\mathbf{g_u}}
\mnewcommand{\gradp}{\mathbf{g}_c}
\mnewcommand{\presh}{p_h}
\mnewcommand{\presc}{p_c}
\mnewcommand{\pres}{p}
\mnewcommand{\nuh}{\mathbf{\nu}_h}
\mnewcommand{\nut}{\nu_{t}}
\mnewcommand{\massf}{m_f}
\mnewcommand{\massfik}{\mathbf{m}_f_{ik}}
\mnewcommand{\bodyforce}{\boldsymbol{f}}

\def\BibTeX{{\rm B\kern-.05em{\sc i\kern-.025em b}\kern-.08em
    T\kern-.1667em\lower.7ex\hbox{E}\kern-.125emX}}
\begin{document}

\title{Performance  assessment of CUDA and OpenACC in large scale combustion simulations}

\author[1]{Guillermo Oyarzun  \thanks{ Corresponding author.guillermo.oyarzun@bsc.es}}
\author[1]{Daniel Mira}
\author[1]{Guillaume Houzeaux}

\affil[1]{Computer Applications in Science and Engineering, Barcelona
Supercomputing Center, Spain}
\maketitle

\begin{abstract}

GPUs have climbed up to the top of supercomputer systems making life harder to many legacy scientific codes. Nowadays, many recipes are being used in such codes' portability, without any clarity of which is the best option. We present a comparative analysis of the two most common approaches, CUDA and OpenACC, into the multi-physics CFD code Alya. Our focus is the combustion problems which are one of the most computing demanding CFD simulations. The most computing-intensive parts of the code were analyzed in detail. New data structures for the matrix assembly step have been created to facilitate a SIMD execution that benefits vectorization in the CPU and stream processing in the GPU. As a result, the CPU code has improved its performance by up to 25\%. In GPU execution, CUDA has proven to be up to $2\times$ faster than OpenACC for the assembly of the matrix.  On the contrary, similar performance has been obtained in the kernels related to vector operations used in the linear solver, where there is minimal memory reuse. 
\end{abstract}


\section{Introduction}

A consolidated trend in the pre-exascale systems is the incorporation of heterogeneous CPU/GPU supercomputers.  One of the main challenges of the scientific community is to adapt their legacy codes to unlock the potential of the modern HPC systems~\cite{donga11}.
 Different programming tools with different levels of complexity have been developed to adopt the GPU computing paradigm. The GPUs require exposing massive parallelism in the algorithms for attaining their maximum performance. This task is far from trivial and requires data restructuring at different levels, according to the computing paradigm utilized. Nowadays, the most common manner to use GPUs is through CUDA or OpenACC.  CUDA~\cite{cuda} is a programming language specifically designed to utilize the NVIDIA GPUs. Its utilization requires more detailed knowledge of the device architecture and a more extensive code refactoring. 
 On the other hand, OpenACC~\cite{openacc} is a directive-based parallel programming model developed to adapt the codes to GPU computing with minimum intervention.  The latter has been widely used during the last years in the portability of real scale scientific applications due to its easiness and straightforward approach. 
 One of the early attempts of using heterogeneous systems for simple CFD simulations was focused on 3D Finite Differences using CUDA for structures meshes and high order schemes~\cite{mic09}. Other examples~\cite{els08,alf11},  of Navier-Stokes simulations on accelerators can be found in the literature. However, they were focused on structured meshes in single GPU implementations. Within the class of CFD algorithms on unstructured meshes, some successful examples\cite{jac13,oya13,ALV18}, based the implementation in an MPI+CUDA execution model, obtaining up to 8 times of acceleration. In the context of reactive flows, the developments have been oriented in the portability of chemical kinetics simulations on GPUs~\cite{nie14,shi11}, using high order methods. Recently, efforts of porting airplane aerodynamics~\cite{fgcs20} and coastal flow simulations~\cite{oya20} using MPI+OpenACC have validated this alternative of computation for large-scale problems. Nevertheless, in each of these examples, the implementation aimed at porting only one physical model without comparing the computational paradigms for using GPUs.
 Our application context is the simulation of real-life combustion systems due to their enormous computing demands. A framework capable of combining advanced turbulence, spray, and combustion models are needed in such simulations. Furthermore, the use of complex geometries and unstructured meshes is essential to work in realistic conditions. Still, it increases substantially the complexity of the discretization strategies and the computational resources to be utilized. This work attempts to show a performance analysis using CUDA and OpenACC tuned versions of a multi-physics CFD code, Alya.  Moreover, general guidelines and common drawbacks when working with legacy codes are going to be presented. Alya is one of the twelve simulation codes of the Unified European Applications Benchmark Suite (UEABS) and thus complies with the highest standards in HPC. The main kernels of the CFD simulation are studied in detail. The numerical tests have been performed in the CTE-Power9~\cite{power9} cluster at the Barcelona Supercomputing Center. 
 The rest of the paper is organized as follows. In the next section, we present the CFD simulations' application context that permits the identification of the most time-consuming parts of the code. In Section~\ref{sec:p9} we present an overview of the hardware architecture and software components of nodes utilized in this work. Section~\ref{sec:portability} describes the general considerations for porting our CFD code to GPU execution. Afterward, in Section~\ref{sec:res}, the impact of the vectorization and the performance of both GPU models are presented. Finally, we summarize our contributions in Section~\ref{sec:conclus}.

\section{Application Context: combustion simulations}
\label{sec:app}
%
\begin{figure*}[ht]
    \centering
    \includegraphics[scale=.65]{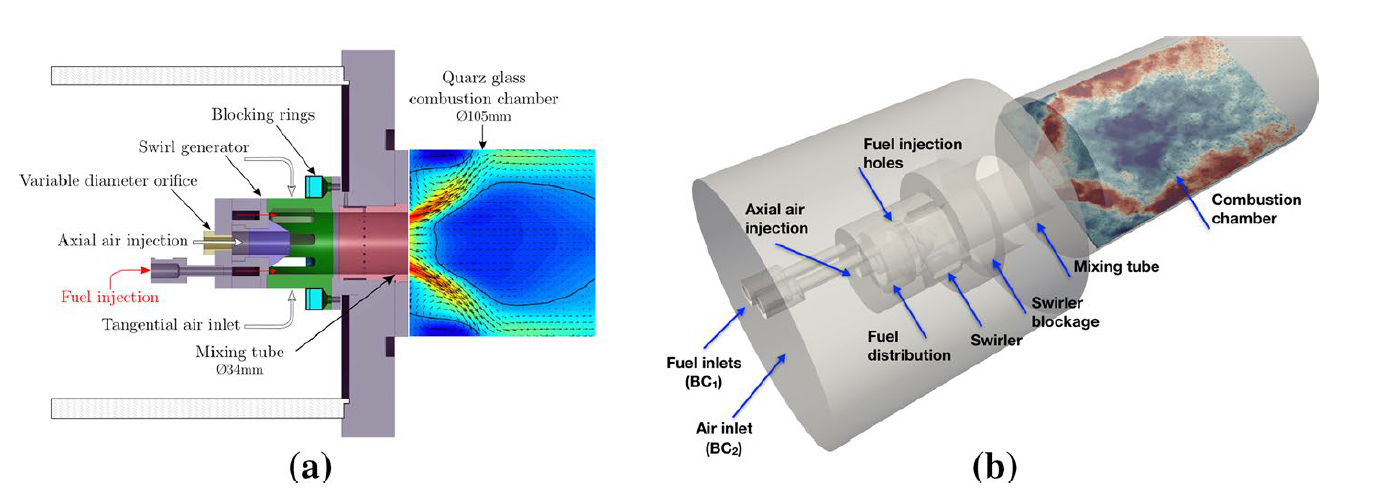}
    \caption{Left. Experimental configuration. Right. Computational domain.}
    \label{fig:domain}
\end{figure*}

The spatially filtered Navier-Stokes equations are considered to model incompressible turbulent flows. 
In short, the problem statement is: find the filtered fluid velocity $\boldu$ and modified pressure $p$ in a domain $\Omega$ and during a given time interval such that:
\begin{EQA}[l]
  \displaystyle{\rho \frac{\partial \boldu}{\partial t}}
   + \rho \left[ 
  2\mathbf{u} \cdot \defor(\mathbf{u})
  +\left(\nabla\cdot\mathbf{u}\right)\mathbf{u}
  -\frac{1}{2} \nabla |\mathbf{u}|^2 \right] \\
   \qquad \qquad \qquad \qquad \qquad \qquad
  - \nabla \cdot[ 2 \mu \defor(\boldu) ]
  + \nabla p + \nabla \cdot \boldtau = \boldo, \label{eq:momentum}\\
   \qquad\qquad\qquad 
   \nabla \cdot \boldu = 0,
   \label{eq:continuity}
\end{EQA}
Where $\rho$ and $\mu$ are the density and viscosity of the fluid, respectively. 

A finite element formulation is used to discretize the computing domain. Note that the mesh used for such a purpose is unstructured and composed of three different kinds of elements (tetrahedra, prisms, and pyramids). We use a fractional step method and a Runge-Kutta scheme of third-order as a time discretization scheme as shown in Algorithm~\ref{alg:kernels}. 
Details on the numerical and time integration schemes are provided in ~\cite{fgcs20}.

\begin{algorithm}[h!]
  \begin{algorithmic}[1]
  \State Element assembly the Laplacian matrix.
  \For   {Time steps}
  \For   {Runge-Kutta steps}  
    \State {\bf  Element assembly}: assembly of momentum equations.  \label{ite:rk1}
    \State {\bf  Boundary assembly}: assembly of wall model.         \label{ite:rk2}
    \State Momentum update.                                          \label{ite:rk3}
    \State Velocity correction.                                      \label{ite:rk4}
  \EndFor
    \State {\bf  Algebraic solver}: solution of pressure equation with the preconditioned CG. \label{ite:press}
  \EndFor
\end{algorithmic}
\caption{Main steps of the fractional step method.}
\label{alg:kernels} 
\end{algorithm} 

The combustion process is assumed to take place in the flamelet regime. The description of the reacting flow in this regime is based on the assumption that chemical time scales are much faster than turbulent time scales, and hence, the thermochemical properties of the flame are described from a precomputed flamelet database. The details on the numerical models used for the Flamelet model are detailed in\cite{mir20}. In short, two extra sets of equations, the heat transfer and the concentration of species, need to be solved each time integration step. 
Such equations are solved using a Runge-Kutta scheme of third-order as in Algorithm~\ref{alg:kernels}, but without needing an algebraic solver. The primary operations within the full algorithm are the matrix assembly, boundary assembly, and the algebraic solver. 
The application context corresponds to a swirl-stabilized technically premixed burner depicted in Fig~\ref{fig:domain}. Further information of the experiment is given in~\cite{tan15,rei17}. In this kind of problem, the most time-consuming operations are the matrix assembly and the algebraic solver. Table~\ref{tab:matass} shows the relative weight of the solver's primary operations for a burner simulation. The chemical model considers two concentration variables, and the algebraic solver uses an iterative preconditioned conjugate gradient. The two main operations sum up to $99.84\%$ of the total execution time and, therefore, become our study's focus.

\begin{table}[htbp]
\caption{Relative weight of the main operations within the time integration step}
\begin{center}
\begin{tabular}{|c|c|c|c|c|}
\hline
\textbf{} & \multicolumn{4}{|c|}{\textbf{Numerical Equations}} \\
\cline{2-5} 
\textbf{Operation} & \textbf{\textit{Navier-Stokes}}& \textbf{\textit{Heat}}& \textbf{\textit{Chemics}}& \textbf{\textit{Total}} \\
\hline
\textbf{Matrix Assembly} & $43.14\%$ & $12.66\%$ & $39.57\%$ & $95.37\%$  \\
\textbf{Boundary Assembly} & $0.03\%$ & $0.02\%$ & $ 0.04\%$  & $0.09\%$ \\
\textbf{Algebraic Solver} & $4.47\%$ & $-$ &  $-$  & $4.47\%$
\\
\textbf{Others} & $0.01\%$ & $0.01\%$ & $ 0.05\%$  & $0.07\%$
\\ \hline
\textbf{Total} & \textbf{$47.65\%$} & \textbf{$12.69\%$} & \textbf{$39.66\%$} & \textbf{$100\%$}  \\
\hline
\end{tabular}
\label{tab:matass}
\end{center}
\end{table}

\section{CTE-Power9 system}
\label{sec:p9}
\begin{figure}[h]
    \centering
    \includegraphics[scale=.45]{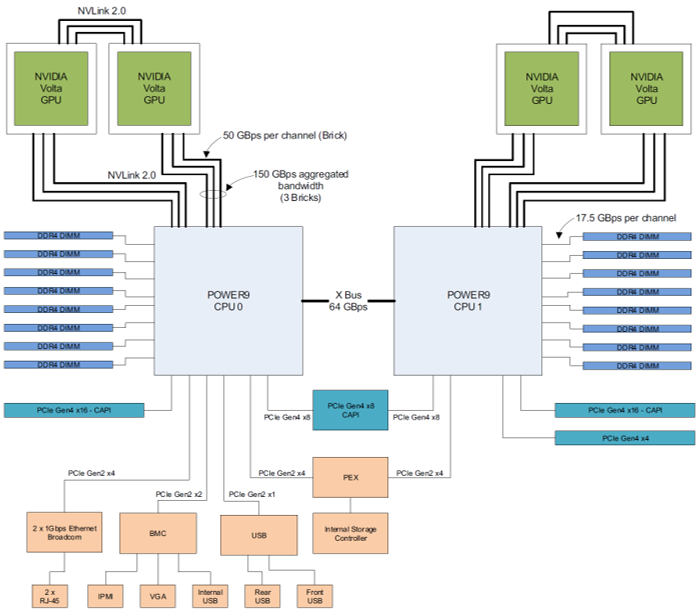}
    \caption{Power 9 cluster configuration.}
    \label{fig:p9}
\end{figure}
Following the current trend in HPC node design, the POWER9 nodes are high throughput nodes (see Fig.~\ref{fig:p9}. 
Each compute node has 2 x POWER9 8335 @3.0 GHz with 20 physical cores each. 
With 4 SMT threads per core, each node can run up to 160 CPU threads. 
The 512 GB of main memory is distributed in 16 DIMMS of 32 GB, each operating at 2666MHz. 
The 4 Volta V100 accelerators have a 7.8 TFLOPS double-precision peak performance giving each one a total of 30.8 TFLOPS. 
The  POWER9 nodes include high throughput NVLINK 2.0 connectors. 
Each GPU has 6 NVLINK connectors, which are attached to the neighboring GPU as well as CPU, 
giving an aggregate bandwidth of 150 GB/s for GPU to GPU as well as GPU to CPU communication.
Details on the node architecture can be found in ~\cite{power9}.

\section{Code Portability}
\label{sec:portability}

\subsection{CUDA vs OpenACC}
CUDA is a parallel computing programming model developed by NVIDIA to utilize its GPUs~\cite{cuda}. 
It provides a set of new instructions used to control the GPU execution and the memory transfers between CPU and GPU. This new instruction set provides more flexibility to the programmer to write GPU code. However, this extra control means more programming complexity. For this reason, the CUDA programmers require to fully understand each detail of hardware and software.

OpenACC offers a different approach to manage GPU computing~\cite{openacc}. Its programming model is based onon compiler directives' utilization to accelerate sections of the code where repetition constructs, such as loops, are found. Only a few compiler directives need to be added to the code, and ideally, without introducing major changes in the data structures. This simplicity comes at expenses on the detailed control flow of the code.
\subsection{Domain discretization}

\begin{figure}[h]
    \centering
    \includegraphics[scale=.3]{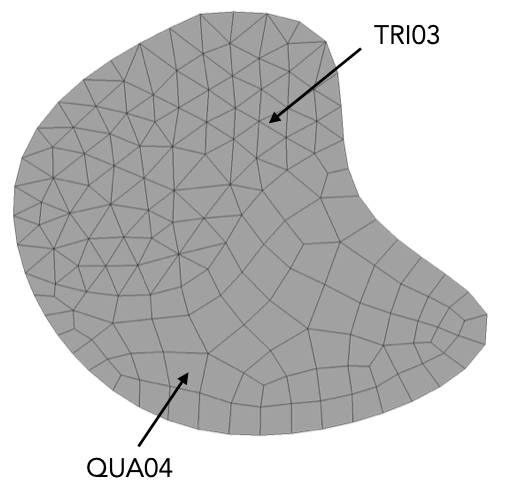}
    \caption{Example of an 2D unstructured mesh with two different geometries.}
    \label{fig:mesh}
\end{figure}

A 3D unstructured mesh is utilized in the discretization of the combustion problems. Commonly, this mesh is composed of different geometrical elements (tetrahedrons, pyramids, and hexahedrons) in which the equations are discretized. An example of a 2D mesh with different kinds of elements is shown in Fig.~\ref{fig:mesh}. 

The matrix assembly and algebraic solver operations are written as for loops that sweep this computational domain. In the matrix assembly, the outer loop moves through the domain's cells (elements), while the inner loop performs the integration on the element. The inner loop is shown in Listing \ref{lst:simple}. The element assembly depends on two variables the number of gauss points (ngauss) and the number of integration nodes (nnodes).  The elemental matrix (Ae) is calculated using the shape functions (N) and the element Jacobian matrix (Jac).

\begin{lstlisting}[ basicstyle={\scriptsize}, caption={Matrix element assembly},label=lst:simple]

do ig = 1,ngauss
  do jn = 1,nnodes
    do in = 1,nnnodes
      Ae(in,jn) = Ae(in,jn) + Jac(ig) * N(in,ig) * N(jn,ig)
    end do
  end do
end do

\end{lstlisting}

In the algebraic solver, the vector operations only perform the outer loop through the domain's cells. The inner loop only exists in the sparse matrix-vector product (SpMV) sparsity pattern depending on the cell's faces. The sparse matrix is stored in the traditional CSR format. Note that all the linear solver operations can be linked with highly optimized libraries like \texttt{cuBLAS} and \texttt{cuSPARSE}.

Different types of elements provoke irregular data paths due to the inner loop's different lengths in the matrix assembly and SpMV.

\subsection{Vectorization}
\begin{figure}[h]
    \centering
    \includegraphics[scale=.25]{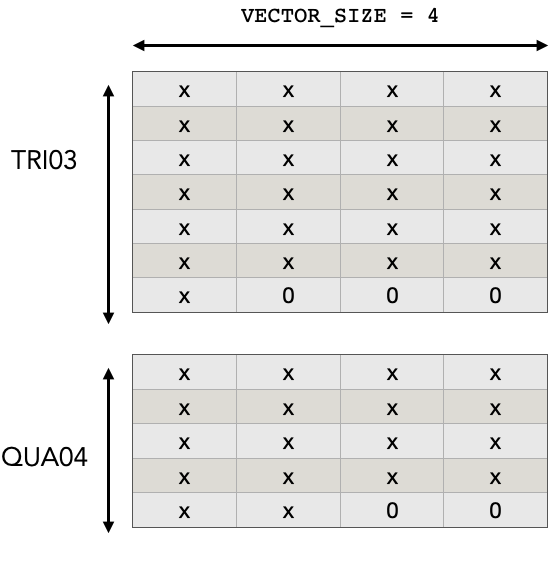}
    \caption{New data structure for a 2D mesh}
    \label{fig:datastruct}
\end{figure}

Nowadays, all CPUs have integrated vector registers that allow performing arithmetic operations in a Single Instruction Multiple Data (SIMD) execution-model. On the other hand, GPUs utilize stream multiprocessors (SM) that execute threads concurrently in an execution model similar to SIMD. Context switching can maintain thousands of threads active and hide the memory latency. Keeping the threads active is known as occupancy, and it is one of the main factors to attain maximum performance in GPU computing.  For this purpose, the kernel workload has to be divided into threads with low register and shared memory requirements. 
Additionally, The creation of new data structures requires special attention. The CPUs' SIMD model and the stream multiprocessors need coalesced and aligned memory accesses for optimal performance. Random memory accesses serialize the execution of the memory fetch and decrease the overall performance.
A data structure that benefits CPU and GPU execution has been developed. First, a renumbering is introduced with the objective of grouping the elements of the same geometrical type. Within each subgroup, the elements are packed in sets of \texttt{VECTOR\_SIZE} elements. If the number of elements in the subgroup is not multiple of \texttt{VECTOR\_SIZE}, then zeros are padded at the end of the pack to maintain regularity. Figure~\ref{fig:datastruct} shows the element organization for a mesh with 25 elements \texttt{TRI03}, 18 elements \texttt{QUAD04} and a \texttt{VECTOR\_SIZE=4}.

The element matrices Ae and Jac are transformed in tensors by adding an extra dimension to the arrays. This new dimension represents the position of the element within the pack. The vectorization consists of performing the matrix assembly for the pack's elements at once, as shown in Listing~\ref{lst:vec}.  Fortran follows a column-major order for storing the arrays. Therefore, the element matrices data is already in a proper order to perform SIMD or SIMT operations. 

\begin{lstlisting}[basicstyle={\scriptsize}, caption={Matrix element assembly VECTOR\_SIZE=4},label=lst:vec]


do ig = 1,ngauss
  do jn = 1,nnodes
    do in = 1,nnnodes
      Ae(1:4,in,jn) = Ae(1:4,in,jn) + &
      Jac(1:4,ig) * N(in,ig) * N(jn,ig)
    end do
  end do
end do

\end{lstlisting}

\section{Numerical Results}
\label{sec:res}
The numerical results have been performed in one node of the Power9 facilities detailed in section~\ref{sec:p9}. The two most time-consuming parts of the Algorithm~\ref{alg:kernels}, the solution of the algebraic solver and the matrix assembly, have been studied using the mesh for the simulation of the burner.

\subsection{Vectorization}

The benefits of vectorization are evident in the matrix assembly. This kernel sums up to $95\%$ of the execution time as shown in Table~\ref{tab:matass}. Our burner mesh consists of 64 million cells with three different elements: tetrahedrons, pyramids, and hexahedrons. In this case, the vectorization improves the matrix assembly's performance in a $25\%$ average. The algebraic solver does not benefit from the new data structure since its algebraic operations were already vectorizing correctly.

\subsection{CUDA vs OpenACC}

\begin{figure}[h]
    \centering
    \includegraphics[scale=.65]{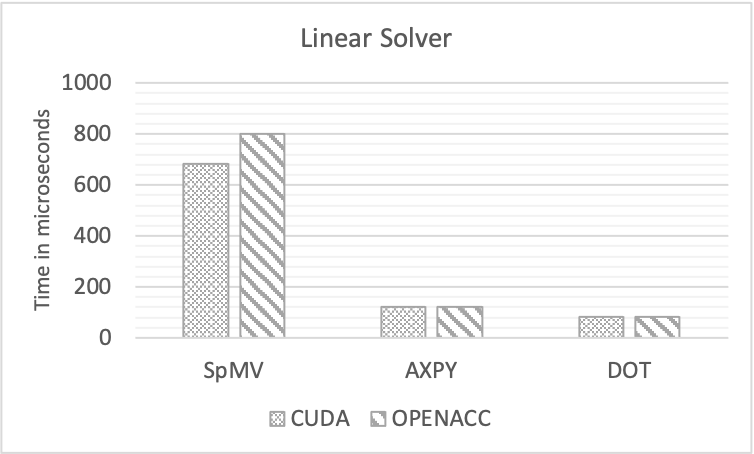}
    \caption{Time in microseconds of the main operations in the algebraic solver.}
    \label{fig:solverres}
\end{figure}

 Our results show two different behavior in both sections of the code. First, the algebraic solver is composed of vector-based operations and the sparse matrix-vector multiplication (SpMV). The vector operations of the algorithm  (AXPY and dot product) do not have memory reuse, and therefore its performance can be easily mimicked by OpenACC. The SpMV, with its indirect memory accesses, is a more complex kernel to replicate by OpenACC, thus behaving slightly worse than CUDA($\approx 10\%$ of slowdown). The results can be seen in Figure \ref{fig:solverres}. 

\begin{figure}[h]
    \centering
    \includegraphics[scale=.65]{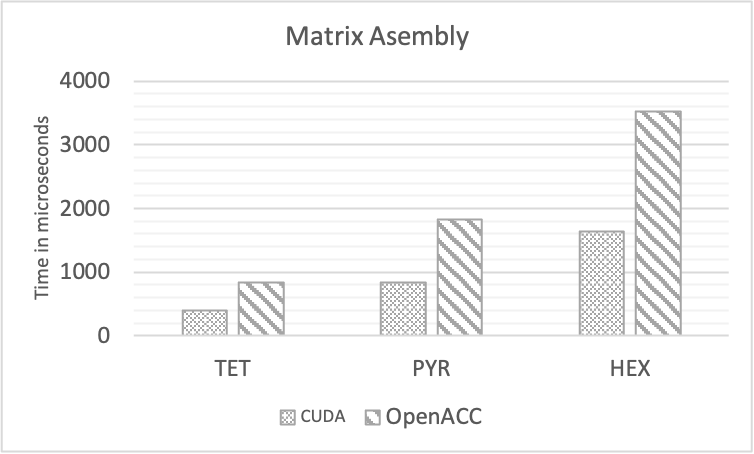}
    \caption{Time in microseconds of the matrix assembly for different type of elements.}
    \label{fig:assemblyres}
\end{figure}

On the other hand, the matrix assembly can be seen as a more complex version of the SpMV, and therefore, the OpenACC performance decreases, reaching up to $2$ times slower than the CUDA version as seen in Figure \ref{fig:assemblyres}. Note that the performance is related to the type of element used in the discretization. The number of nodes used in the discretization determines the memory pattern and footprint. For the same version of the CUDA kernel, the performance decreases as the number of nodes are higher. More complex CUDA kernels need to be developed specifically for the type of element, losing in the generality and reusability of the code to maintain the performance.

\section{Conclusions}
\label{sec:conclus}

This work has shown the benefits of using new data structures that exploit the SIMD model execution in CPUs and GPUs. In the CPUs, the code can improve its performance $25\%$ on average. In GPUs, the new data structures expose the data parallelism and facilitate OpenACC or CUDA kernels' implementation.  Specific CUDA kernels can outperform up to $2$ times the OpenACC implementation in the matrix assembly operation. However, this improvement comes at the expense of many programming hours. For the solver's algebraic operations, OpenACC seems like the best option since a similar performance can be achieved but without increasing the complexity of the code.

\section*{Acknowledgments}
This work is partially supported by the European Union's Horizon 2020 research and innovation programme under grant agreement number: 846139 (Exa-FireFlows). It is also partially supported by the BSC-IBM Deep Learning Research Agreement, under JSA “Application porting, analysis and optimization for POWER and POWER AI”. It has also received funding by the EXCELLERAT project funded by the European Commission’s ICT activity of the H2020 Programme under grant agreement number: 823691.
This paper expresses the opinions of the authors and not necessarily those of the European Commission. The European Commission is not liable for any use that may be made of the information contained in this paper.

\end{document}